\documentclass[11pt,titlepage,a4paper]{article}
\usepackage{bozhomac}	
\usepackage{bozhlogo}

\title{\bfseries	\vspace*{-2.302345in}
\vspace*{-3ex}
{
\begin{flushright}
	  \textbf{\large LANL xxx E-print archive No. quant-ph/9803084}\\[2ex]
\end{flushright}
}
{\huge Fibre bundle formulation of \\[6pt]
nonrelativistic quantum mechanics  \\[2ex]
\Large I. Introduction. The evolution transport
}
}

\vspace{2ex}

\author{
Bozhidar Z. Iliev
\thanks{Department Mathematical Modeling,
Institute for Nuclear Research and \mbox{Nuclear} Energy,
Bulgarian Academy of Sciences,
Boul. Tzarigradsko chauss\'ee~72, 1784 Sofia, Bulgaria}
\thanks{E-mail address: bozho@inrne.bas.bg}
\thanks{URL: http://www.inrne.bas.bg/mathmod/bozhome/}
}

\date{
\vspace{2.27ex}\ShortTitle{Bundle quantum mechanics: I}	\\[0.27ex]
\vspace{3.27ex}
	\begin{tabular}{r@{$\colon\to~$}l}
\vspace{0.09ex} Basic ideas	& March 1996		\\[0.09ex]
\vspace{0.09ex} Began		& May 19, 1996		\\[0.09ex]
\vspace{0.09ex} Ended		& July 12, 1996		\\[0.09ex]
\vspace{0.09ex} Revised		& December 1996 -- January 1997,\\[0.09ex]
\vspace{0.09ex} Revised		& April 1997, September 1998	\\[0.09ex]
\vspace{0.09ex} Last update    	  & October 15, 1998  	\\[0.09ex]
\vspace{0.09ex} Composing part I  & September 23, 1997	\\[0.09ex]
\vspace{0.09ex} Extracting part I & October 4, 1997	\\[0.09ex]
\vspace{0.09ex} Updating part I   & October 16, 1998	\\[0.09ex]
\vspace{0.27ex} Produced	& \fbox{\today}		\\[0.27ex]
	\end{tabular} \\[1.27ex]
	\begin{tabular}{r@{$\colon~$}l}
\vspace{0.27ex} LANL xxx archive server E-print No.& quant-ph/9803084
							\\[0.27ex]
	\end{tabular} \\[-0.27ex]
\vspace{3.0ex}{\Huge\BOZHO}	\\[3.0ex]
\vspace{0.27ex}\Subject{Quantum mechanics; Differential geometry} \\[2.27ex]
	\begin{tabular}{r@{\hspace{0.512em}}|@{\hspace{0.512em}}l}
\vspace{0.27ex}\MSC[1991]{81P05, 81P99, 81Q99, 81S99}		  
&
\vspace{0.27ex}\PACS[1996]{02.40.Ma, 04.60.-m, 03.65.Ca, 03.65.Bz}
	\end{tabular} \\[1.27ex]
\vspace{0.27ex}\KeyWords{Quantum mechanics; Geometrization of quantum
		mechanics;\\ Fibre bundles}	\\[0.27ex]
}
\newcommand{\Hil}{\mathcal{F}}	
\newcommand{\HilB}{(\bHil,\proj,\base)}	
	\newcommand{\bHil}{\mathit{F}}	
	\newcommand{\proj}{\pi}		
	\newcommand{\base}{\mathit{M}}	

\newcommand{\Ham}{\mathcal{H}}	

\newcommand{\dyn}[1]{\pmb{\mathbb{#1}}}	
	\newcommand{\ope}[1]{\mathcal{#1}}		 
	\newcommand{\mor}[1]{\mathit{#1}}		 

\providecommand{\iu}{\mathrm{i}} 
\newcommand{\ih}{\mathrm{i}\hbar}
\newcommand{\iih}{\frac{1}{\ih}} 

\begin{document}		

\renewcommand{\thefootnote}{\fnsymbol{footnote}}
\maketitle			
\renewcommand{\thefootnote}{\arabic{footnote}}

\tableofcontents		


\pagestyle{myheadings}
\markright{\itshape\bfseries Bozhidar Z. Iliev:
	\upshape\sffamily\bfseries Bundle quantum mechanics.~I}

\begin{abstract}

We propose a new systematic fibre bundle formulation of nonrelativistic
quantum mechanics. The new form of the theory is equivalent to the usual one
but it is in harmony with the modern trends in theoretical physics and
potentially admits new generalizations in different directions. In it
a pure state of some quantum system is described by a state section (along
paths) of a (Hilbert) fibre bundle. Its evolution is determined through the
bundle (analogue of the) Schr\"odinger equation. Now the dynamical variables
and the density operator are described via bundle morphisms (along paths).
The mentioned quantities are connected by a number of relations derived in
this work.
%
%

	The present first part of this investigation is devoted to the
introduction of basic concepts on which the fibre bundle approach to quantum
mechanics rests. We show that the evolution of pure quantum-mechanical states
can be described as a suitable linear transport along paths, called evolution
transport, of the state sections in the Hilbert fibre bundle of states of a
considered quantum system.

\end{abstract}

\section {Introduction}
\label{introduction-I}
\setcounter{equation} {0}

	Usually the standard nonrelativistic quantum mechanics of pure states
is formulated in terms of vectors and operators in a Hilbert
space~\cite{Dirac-PQM, Fock-FQM, Messiah-QM, Prugovecki-QMinHS, L&L-3}.
This is in discrepancy and not in harmony with the new trends in
(mathematical) physics~\cite{Schutz, Coquereaux, Konopleva-Popov}
in which the theory of fibre
bundles~\cite{Husemoller, R_Hermann/Geom-phys-systems},
in particular vector
bundles~\cite{R_Hermann-I, R_Hermann-II},
is essentially used. This paper (and its further continuation(s)) is intended
to incorporate the quantum theory in the family of fundamental physical
theories based on the background of fibre bundles.

	The idea of geometrization of quantum mechanics is an old one (see,
e.g.,~\cite{Kibble-79} and the references therein). A good motivation for
such approach is given in~\cite{Anandan-90a,Kibble-79}.
Different geometrical structures in quantum mechanics were
introduced~\cite{Ashtekar&Schilling,Brody&Hughston},
for instance such as inner
products(s)~\cite{Fock-FQM,Messiah-QM,Kibble-79,Anandan-90b},
(linear) connection~\cite{Anandan-90a,Anandan-90b,Uhlmann-91a},
symplectic structure~\cite{Anandan-90a},
complex structure~\cite{Kibble-79}, etc.
The introduction of such structures admits a geometrical treatment of some
problems, for instance, the dynamics in the (quantum) phase
space~\cite{Kibble-79}
and the geometrical phase~\cite{Anandan-90a}.
In a very special case a gauge structure, i.e.\  a parallel transport
corresponding to a linear connection, in quantum mechanics is
pointed out in~\cite{Wilczek-Zee}.
For us this work is remarkable with the fact that the equation~(10) found
in it is a very `ancient' special version of
%
%
the transformation law for the matrix-bundle Hamiltonian, derived in this
investigation, which, together with the bundle (analog of the) Schr\"odinger
equation,
shows that (up to a constant) with respect to the
quantum evolution the Hamiltonian plays the r\^ole of a gauge field
(connection). In~\cite{Uhlmann-91a,Uhlmann-91b} one finds different (vector)
bundles defined on the base of the (usual) Hilbert space of quantum mechanics
or its modifications. In these works different parallel
transports in the corresponding bundles are introduced too.

	A general feature of all of the references above-cited is that in
them all geometric concepts are introduced by using in one or the other way
the accepted mathematical foundation of quantum mechanics, viz.\  a
suitable Hilbert or projective Hilbert
space and operators acting in it. The Hilbert space may be
extended in a certain sense or replaced by a more general space, but this
does not change the main ideas. One of the aims of this work is namely to
change this mathematical background of quantum mechanics.

	Separately we have to mention the approach of Prugove\v{c}ki to the
quantum theory, a selective summary of which can be found
in~\cite{Drechsler/Tuckey-96} (see also the references therein) and
in~\cite{Coleman-96}. It can be characterized as `stochastic' and
`bundle'. The former feature will not be discussed in the present
investigation; thus we lose some advantages of the stochastic quantum theory
to which we shall return elsewhere. The latter `part' of the Prugove\v{c}ki's
approach has some common aspects with our present work but, generally, it is
essentially different. For instance, in both cases the quantum evolution from
point to point (in space-time) is described via a kind of (parallel or
generic linear) transport (along paths) in a suitable Hilbert fibre bundle.
But the notion of a `Hilbert bundle' in our and Prugove\v{c}ki's approach is
different nevertheless that in both case the typical (standard) fibre is
practically the same (when one and the same theory is concerned). Besides, we
need not even to introduce the Poincar\'e (principal) fibre bundle over the
space-time or the phase space which play an important r\^ole in
Prugove\v{c}ki's theory. Also we have to notice that the used in it concepts
of quantum and parallel transport are special cases of the notion of a
`linear transport along paths' introduced
in~\cite{bp-normalF-LTP,bp-LTP-general}.
The application of the last concept, which is accepted in the present
investigation, has a lot of advantages, significantly simplifies some proofs
and makes certain results `evident' or trivial (e.g. the last part of
section~2 and the whole section~4 of~\cite{Drechsler/Tuckey-96}). At last, at
the present level (nonrelativistic quantum mechanics) our bundle formulation
of the quantum theory is insensitive with respect to the space-time
curvature. A detail comparison of Prugove\v{c}ki's and our approaches to the
quantum theory will be done elsewhere.

	Another geometric approach to quantum mechanics is proposed
in~\cite{Graudenz-94} and partially in~\cite{Graudenz-96}, the letter of
which is, with a few exceptions, almost a review of the former.
These works suggest two ideas which are quite important for us.
First, the quantum evolution could be described as a (kind of) parallel
transport in an infinitely dimensional (Hilbert) fibre bundle over the
space-time. And second, the concrete description of a quantum system should
explicitly depend on (the state of) the observer with respect to which it is
depicted (or who `investigates' it). These ideas are incorporated and
developed in our work.

	From the known to the author literature, the closest to the
approach developed in this work is~\cite{Asorey&Carinena&Paramio} which
contains an excellent motivation for applying the fibre bundle technique to
nonrelativistic quantum mechanics.%
\footnote{%
The author thanks J.F.~Cori\~nena (University of Zaragoza, Zaragoza, Spain)
for drawing his attention to reference~\cite{Asorey&Carinena&Paramio}
in May 1998.%
}
Generally said, in this paper the evolution of a quantum system is described
as a `generalized parallel transport' of appropriate objects in a Hilbert
fibre bundle over the 1\ndash dimensional manifold
$\mathbb{R}_+:=\{t :  t\in\mathbb{R}, t\ge0\}$,
interpreted as a `time' manifold (space). We shall comment on
reference~\cite{Asorey&Carinena&Paramio} in the second part of this series,
after developing the formalism required for its analysis. Besides, the
paper~\cite{Asorey&Carinena&Paramio} contains an excellent motivation for
applying the apparatus of fibre bundle theory to quantum mechanics.

	An attempt to formulate quantum mechanics in terms of a fibre bundle
over the phase space is made in~\cite{Reuter}. Regardless of some common
features, this paper is quite different from the present investigation.
We shall comment on it later.
In particular, in~\cite{Reuter} the gauge structure of the arising theory is
governed by a non\ndash dynamical connection related to the symplectic
structure of the system's phase space, while in this work analogous structure
(linear transport along paths) is uniquely connected with system's
Hamiltonian, playing here the r\^{o}le of a gauge field itself.

	The present work is a direct continuation of the considerations
in~\cite{bp-BQM-preliminary} which paper, in fact, may be regarded as its
preliminary  version. Here we suggests a purely fibre bundle formulation of
the nonrelativistic quantum mechanics. This new form of the theory is
entirely equivalent to the usual one, which is a consequence or our step by
step equivalent reformulation of the quantum theory. The bundle description
is obtained on the base of the developed by the author theory of transports
along paths in fibre
bundles~\cite{bp-normalF-LTP,bp-TP-general,bp-LTP-general}, generalizing the
theory of parallel transport, which is partially generalized here to the
infinitely dimensional case.

        The main object in quantum mechanics is the Hamiltonian (operator)
which, through the Schr\"odinger equation, governs the evolution of a quantum
system~\cite{Messiah-QM,Fock-FQM, Prugovecki-QMinHS, L&L-3}. In our novel
approach its r\^ole is played by a suitable linear transport along paths
in an appropriate (Hilbert) fibre bundle. It turns out that up to a
constant the matrix-bundle Hamiltonian, which is uniquely
determined by the Hamiltonian in a given field of bases, coincides with
the matrix of the coefficients of this transport
(cf. an analogous result in~\cite[sect.~5]{bp-BQM-preliminary}).
This fact, together with the replacement of the usual Hilbert space with a
Hilbert fibre bundle, is the corner-stone for the possibility for the new
formulation of the nonrelativistic quantum mechanics.

\vspace{2ex}
	The present first part of our investigation is organized as follows.

        In Sect.~\ref{II} are reviewed some facts from the quantum mechanics
and partially is fixed our notation. Here, as well as throughout this
work, we follow the established in the physical literature degree of rigor.
But, if required, the present work can reformulate to meet the
present-day mathematical standards. For this purpose one can use, for
instance, the quantum-mechanical formalism described
in~\cite{Prugovecki-QMinHS} or in~\cite{Neumann-MFQM}
(see also~\cite{Reed&Simon}).

        In Sect.~\ref{III} we recall the notion of a linear transport along
paths in vector fibre bundles and make certain remarks concerning the
special case of a Hilbert bundle.

	Sect.~\ref{new-I} begins the building of the new bundle approach to
quantum mechanics. Here the concept of a \emph{Hilbert fibre bundle of the
states} corresponding to a quantum system is introduced. The analogue of the
state vector now is the \emph{state section (along paths)}. We present here
also some technical (mathematical) details, such as ones concerning
(Hermitian) bundle metric, Hermitian and unitary maps etc.

	In Sect.~\ref{IV} is proved that in the new description the evolution
operator of a quantum system is (equivalently) replaced by a suitable linear
transport along paths, called
\emph{evolution transport} or a \emph{bundle evolution operator}.

	The paper closes with Sect.~\ref{conclusion-I}.

\section {Evolution of pure quantum states (review)}
\label{II}
\setcounter{equation} {0}

	In quantum mechanics~\cite{Messiah-QM,Fock-FQM,L&L-3} a pure state of
a quantum system is described by a state vector $\psi(t)$ (in
Dirac's~\cite{Dirac-PQM} notation $|t\rangle$) generally depending on the
time $t\in\mathbb R$ and belonging to a Hilbert space $\Hil$ (specific to any
concrete system) endowed with a nondegenerate Hermitian scalar product
\(
\langle\cdot | \cdot \rangle\colon  \Hil\times\Hil \to \mathbb{C}.
\)%
\footnote{%
For some (\eg unbounded) states the system's state vectors form a more
general space than a Hilbert one. This is insignificant for the following
presentation.%
}
For any two instants of time $t_2$ and $t_1$ the corresponding state vectors
are connected by the equality
\begin{equation}	\label{2.1}
\psi(t_2) = \ope{U}(t_2,t_1)\psi(t_1)
\end{equation}
where $\ope{U}$ is the \emph{evolution operator} of the
system~\cite[chapter~IV, Sect.~3.2]{Prugovecki-QMinHS}.  It is supposed to be
linear and unitary, i.e.\
	\begin{align}   \label{2.20}
\ope{U}(t_2,t_1)( \lambda\psi(t_1) + \mu\xi(t_1) )	&=
\lambda \ope{U}(t_2,t_1)(\psi(t_1)) + \mu \ope{U}(t_2,t_1)( \xi(t_1) ),
\ \ \ \ \ \ \ \         \\  \label{2.2}
\ope{U}^\dag(t_1,t_2) &= \ope{U}^{-1}(t_2,t_1),
	\end{align}
for any $\lambda,\mu\in\mathbb{C}$ and state vectors $\psi(t),\xi(t)\in\Hil$,
 and such that for any $t$
	\begin{equation}	\label{2.3}
\ope{U}(t,t) = \id_\Hil.
	\end{equation}
Here $\id_X$ means the identity map of a set $X$ and the dagger ($\dag$)
 denotes  Hermitian conjugation, i.e.\  if $\varphi,\psi\in\Hil$ and
$\ope{A}\colon \Hil\to\Hil$, then $\ope{A}^\dag$ is defined by
	\begin{equation}	\label{2.4}
\langle \ope{A}^\dag\varphi | \psi \rangle =
			\langle \varphi | \ope{A}\psi \rangle.
	\end{equation}
In particular $\ope{U}^\dag$ is defined by
\(
\langle \ope{U}^\dag(t_1,t_2)\varphi(t_2) | \psi(t_1) \rangle =
\langle \varphi(t_2) | \ope{U}(t_2,t_1)\psi(t_1) \rangle.
\)
So, knowing $\psi(t_0)=\psi_0$ for some moment~$t_0$, one knows the state
vector for any moment $t$ as
\(
\psi(t)=\ope{U}(t,t_0)\psi(t_0)=\ope{U}(t,t_0)\psi_0.
\)

	Let $\Ham(t)$ be the Hamiltonian (function) of the
system, i.e.\  its total energy operator. It generally depends on the
time~$t$ explicitly%
\footnote{%
Of course, the Hamiltonian depends also on the observer with respect to which
the evolution of the quantum system is described. This dependence is
usually implicitly assumed and not written
explicitly~\cite{Dirac-PQM,Messiah-QM}. This deficiency will be eliminated in
a natural way further in the present work. The Hamiltonian can also depend on
other quantities, such as the (operators of the) system's generalized
coordinates. This possible dependence is insignificant for our investigation
and will not be written explicitly.%
}
and it is supposed to be a Hermitian operator, i.e.\  $\Ham^\dag(t)=\Ham(t)$.
The Schr\"odinger equation
(see~\cite[\S~27]{Dirac-PQM} or~\cite[chapter~V, Sec.~3.1]{Prugovecki-QMinHS})
	\begin{equation}	\label{2.5}
\ih\frac{d \psi(t)}{d t} = \Ham(t) \psi(t),
	\end{equation}
with
$\iu\in\mathbb{C}$ and $\hbar$ being respectively the imaginary unit
and the Plank's constant (divided by $2\pi$),
together with some initial condition
	\begin{equation}	\label{2.5a}
\psi(t_0)=\psi_0\in\Hil
	\end{equation}
is postulated in the quantum mechanics

		The substitution of~(\ref{2.1}) into~(\ref{2.5}) shows that
there is a 1:1 correspondence between~$\ope{U}$ and~$\Ham$ described by
	\begin{equation}	\label{2.6}
\ih\frac{\partial \ope{U}(t,t_0)}{\partial t} = \Ham(t)\circ\ope{U}(t,t_0),
\qquad
\ope{U}(t_0,t_0) = \id_\Hil
	\end{equation}
where $\circ$ denotes composition of maps.
If~$\ope{U}$ is given, then
	\begin{equation}	\label{2.7}
\Ham(t) =
\ih\frac{\partial \ope{U}(t,t_0)}{\partial t} \circ \ope{U}^{-1}(t,t_0) =
\ih\frac{\partial \ope{U}(t,t_0)}{\partial t} \circ \ope{U}(t_0,t),
	\end{equation}
where we have used the equality
\[
\ope{U}^{-1}(t_2,t_1) = \ope{U}(t_1,t_2)
\]
which follows from~(\ref{2.1})
(see also below~(\ref{2.8}) or Sect.~\ref{IV}). Conversely, if
$\Ham$ is given, then~\cite[chapter~VIII, \S\ 8]{Messiah-QM} $\ope{U}$
is the unique solution of the integral equation
\(
\ope{U}(t,t_0) = \id_\Hil +
{\iih} \int\limits_{t_0}^{t} \Ham(\tau)\ope{U}(\tau,t_0) d\tau,
\)
i.e.\  we have
	\begin{equation}	\label{2.8}
 \ope{U}(t,t_0) = \Texp\int\limits_{t_0}^{t}
{\iih} \Ham(\tau) d\tau,
	\end{equation}
where $\Texp\int\limits_{t_0}^{t}\cdots d\tau$ is the chronological
(called also T-ordered, P-ordered or path-ordered) exponent (defined, e.g as
the unique solution of the initial-value problem~(\ref{2.6}); see
also~\cite[equation~(1.3)]{Asorey&Carinena&Paramio}).%
\footnote{%
The physical meaning of $\ope{U}$ as a propagation function, as well as its
explicit calculation (in component form) via $\Ham$ can be found, e.g.,
in~\cite[\S~21, \S~22]{Bjorken&Drell-1}%
}
From here
follows that the Hermiticity of $\Ham$, $\Ham^\dag=\Ham$, is equivalent to
the unitarity of $\ope{U}$ (see~(\ref{2.2})).

	Let us note that for mathematically rigorous understanding of the
derivations in~(\ref{2.5}), (\ref{2.6}), and~(\ref{2.7}), as well as of the
chronological (path-ordered) exponent in~(\ref{2.8}), one has to apply the
developed in~\cite{Prugovecki-QMinHS} mathematical apparatus, but this is out
of the subject of the present work.

	If $\ope{A}(t)\colon \Hil\to\Hil$ is the (linear) operator
corresponding to a dynamical variable
$\dyn{A}$ at the moment $t$, then the mean value
(= the mathematical expectation) which it assumes at a state described by a
state vector $\psi(t)$ with a finite norm is
	\begin{equation}	\label{2.9}
\langle\ope{A}(t)\rangle_\psi^t :=
\frac{\langle\psi(t) | \ope{A}(t)\psi(t)\rangle}
{\langle\psi(t) | \psi(t)\rangle}.
	\end{equation}

	Often the operator $\ope{A}$ can be chosen to be independent
of the time $t$. (This is possible, e.g., if $\ope{A}$ does not depend on $t$
explicitly~\cite[chapter~VII, \S~9]{Messiah-QM} or if the spectrum of
$\ope{A}$ does not change in time~\cite[chapter~III, sect.~13]{Fock-FQM}.) If
this is the case, it is said that the system's evolution is depicted in the
Schr\"odinger picture of motion~\cite[\S~28]{Dirac-PQM}, \cite[chapter~VII,
\S~9]{Messiah-QM}.

\section[Linear transports along paths and Hilbert fibre bundles]
	{Linear transports along paths and \\Hilbert fibre bundles}
\label{III}
\setcounter{equation} {0}

	The general theory of linear transports along paths in vector bundles
is developed at length in~\cite{bp-normalF-LTP,bp-LTP-general}. In the
present investigation we shall need only a few definitions and results from
these papers when the bundle considered is a Hilbert one (see below
definition~\ref{Defn3.2}). To their partial introduction and motivation is
devoted the current section.

	Let $(E,\pi,B)$ be a complex%
\footnote{%
All of our definitions and results hold also for real vector bundles. Most of
them are valid for vector bundles over more general fields too but this is
inessential for the following.%
}
vector bundle~\cite{Greub&et.al.-1} with bundle (total) space $E$, base $B$,
projection $\pi\colon E\to B$, and isomorphic fibres $\pi^{-1}(x)\subset E$,
$x\in B$.  Let $\mathcal{E}$ be the (standard, typical) fibre of the bundle,
\ie a vector space to which all $\pi^{-1}(x)$, $x\in B$ are homeomorphic
(isomorphic).  By $J$ and $\gamma\colon J\to B$ we denote, respectively, a
real interval and path in $B$.

	\begin{Defn}	\label{Defn3.1}
	A linear transport along paths in the  bundle $(E,\pi,B)$ is a map $L$
assigning to any path $\gamma$ a map $L^\gamma$, transport along $\gamma$,
such that $L^\gamma\colon (s,t)\mapsto L^\gamma_{s\to t}$ where the map
	\begin{equation}				   \label{3.fibre}
L^\gamma_{s\to t} \colon  \pi^{-1}(\gamma(s)) \to \pi^{-1}(\gamma(t))
	\qquad s,t\in J,
	\end{equation}
called transport along $\gamma$ from $s$ to $t$, has the properties:
	\begin{alignat}{2}				    \label{3.1}
L^\gamma_{s\to t}\circ L^\gamma_{r\to s} &=
			L^\gamma_{r\to t},&\qquad  r,s,t&\in J, \\
L^\gamma_{s\to s} &= \id_{\pi^{-1}(\gamma(s))}, & s&\in J,   \label{3.2}
\\
L^\gamma_{s\to t}(\lambda u + \mu v) 			     \label{3.linear}
  &= \lambda L^\gamma_{s\to t}u + \mu L^\gamma_{s\to t}v,
	& \lambda,\mu &\in \mathbb{C},\quad u,v\in{\pi^{-1}(\gamma(s))},
	\end{alignat}
where  $\circ$ denotes composition of maps and
$\id_X$ is the identity map of a set $X$.
	\end{Defn}

	\begin{Rem}
	Equations~\eref{3.1} and~\eref{3.2} mean that $L$ is a
\emph{transport along paths} in the bundle
$(E,\pi,B)$~\cite[definition~2.1]{bp-TP-general}, while~\eref{3.linear}
specifies that it is \emph{linear}~\cite[equation~(2.8)]{bp-TP-general}. In
the present paper only linear transports will be used.
	\end{Rem}

	This definition generalizes the concept of a parallel transport in
the theory of (linear) connections (see~\cite{bp-TP-general,bp-TP-parallelT}
and the references therein for details and comparison).

	A few comments on definition~\ref{Defn3.1} are now in order.
According to equation~\eref{3.fibre}, a linear transport along paths may be
considered as a path-depending connection: it establishes a fibre (isomorphic
- see below) correspondence between the fibres over the path along which it
acts. By virtue of equation~\eref{3.linear} this correspondence is linear.
Such a condition is a natural one when vector bundles are involved, it simply
represents a compatibility condition with the vectorial structure of the
bundle (see~\cite[sect.~2.3]{bp-TP-general} for details). Equation~\eref{3.2}
is a formal realization of our intuitive and na\"{\i}ve understanding that if
we `stand' at some point of a path without `moving' along it, then `nothing'
must happen with the fibre over that point. This property fixes a
0\ndash{ary} operation in the set of (linear) transports along paths,
defining in it the `unit' transport. At last, the equality~\eref{3.1}, which
may be called a group property of the (linear) transports along paths, is a
rigorous expression of the intuitive representation that the `composition'
of two (linear) transports along one and the same path must be a (linear)
transport along the same path.

	In general, different forms of~\eref{3.fibre}--\eref{3.linear} are
well know properties of the parallel transports generated by (linear)
connections (see~\cite{bp-TP-parallelT}). By this reason these transports
turn to be special cases of the general (linear) transport along
paths~\cite[theorem~3.1]{bp-TP-parallelT}. In particular,
	comparing definition~\ref{Defn3.1}
with~\cite[definition~2.1]{bp-LT-Deriv-tensors} and taking into
account~\cite[proposition~4.1]{bp-LT-Deriv-tensors}, we conclude that special
types of linear transports along paths are: the parallel transport assigned to
a linear connection (covariant differentiation) of the tensor algebra of a
manifold~\cite{K&N-1,Schouten/Ricci},
Fermi-Walker transport~\cite{Hawking&Ellis,
Synge}, Fermi transport~\cite{Synge},
Truesdell transport~\cite{Walwadkar,Walwadkar&Virkar},
Jaumann transport~\cite{Radhakrishna&et.al.},
Lie transport~\cite{Hawking&Ellis,Schouten/Ricci},
the modified Fermi\ndash Walker and Frenet\ndash Serret
transports~\cite{Dandoloff&Zakrzewski},
etc.
Consequently definition~\ref{Defn3.1} is general enough to cover a list of
important transports used in theoretical physics and mathematics. Thus
studying  the properties of the linear transports along paths we can make
corresponding conclusions for any one of the transports mentioned.%
\footnote{%
The concept of linear transport along paths in vector bundles can be
generalized to the transports along paths in arbitrary
bundles~\cite{bp-TP-general} and to transports along maps in
bundles~\cite{bp-TM-general}. An interesting considerations of the concept of
(parallel) `transport' (along closed paths) in connection with homotopy
theory and the classification problem of bundles can be found
in~\cite{Stasheff-PT}. These generalizations will not be used in the
present work.%
}

	From~\eref{3.1} and~\eref{3.2} we get that
$L^\gamma_{s\to t}$ are invertible and
	\begin{equation}	\label{3.3}
\left(L^\gamma_{s\to t}\right)^{-1} = L^\gamma_{t\to s},
	\qquad s,t\in J.
	\end{equation}
Hence the linear transports along paths are in fact linear isomorphisms of the
fibres over the path along which they act.

	The following two propositions establish the general structure of
linear transports along paths.

	\begin{Prop}	\label{Prop3.1}
	A map~\eref{3.fibre} is a linear transport along $\gamma$ from $s$ to
$t$ for every $s,t\in J$ if and only if there exist an isomorphic with
$\pi^{-1}(x),\ x\in B$ vector space $V$ and family of linear isomorphisms
$\{F(s;\gamma)\colon \pi^{-1}(\gamma(s))\to V,\ s\in J\}$ such that
	\begin{equation}	\label{3.4}
L_{s\to t}^{\gamma} =
	F^{-1}(t;\gamma) \circ  F(s;\gamma),\qquad s,t\in J.
	\end{equation}
	\end{Prop}

	\begin{Proof}
	If~\eref{3.fibre} is a linear transport along $\gamma$ from $s$ to
$t$, then fixing some $s_0\in J$ and using~\eref{3.2} and~\eref{3.3}, we get
\(
L_{s\to t}^{\gamma} = L_{s_0\to t}^{\gamma} \circ L_{s\to s_0}^{\gamma}
	= \bigl(L_{t\to s_0}^{\gamma}\bigr)^{-1} \circ L_{s\to s_0}^{\gamma}.
\)
So~\eref{3.4} holds for $V=\pi^{-1}(\gamma(s_0))$ and
$F(s;\gamma)=L_{s\to s_0}^{\gamma}$. Conversely, if~\eref{3.4} is valid for
some linear isomorphisms $F(s;\gamma)$, then a straightforward calculation
shows that it converts~\eref{3.1} and~\eref{3.2} into identities
and~\eref{3.linear} holds due to the linearity of $F(s;\gamma)$.
	\end{Proof}

	\begin{Prop}	\label{Prop3.2}
	Let in the vector bundle $(E,\pi,B)$ be given linear transport
along paths with a representation~\eref{3.4} for some vector space $V$
and linear isomorphisms $F(s;\gamma)\colon \pi^{-1}(\gamma(s))\to V,\ s\in J$.
Then for a vector space $\lindex[V]{}{\star}$ there exist linear
isomorphisms
\(
\lindex[\mspace{-2mu}F]{}{\star}(s;\gamma)\colon \pi^{-1}(\gamma(s))\to
	\lindex[V]{}{\star},
\ s\in J
\)
for which
	\begin{equation}	\label{3.6}
L_{s\to t}^{\gamma} =
	\lindex[\mspace{-2mu}F]{}{\star}^{-1}(t;\gamma) \circ
	\lindex[\mspace{-2mu}F]{}{\star}(s;\gamma),\qquad s,t\in J.
	\end{equation}
iff there exists a linear isomorphism
$D(\gamma)\colon V\to\lindex[V]{}{\star}$ such that
	\begin{equation}	\label{3.7}
\lindex[\mspace{-2mu}F]{}{\star}(s;\gamma) = D(\gamma)\circ F(s;\gamma),
				\qquad s\in J.
	\end{equation}
	\end{Prop}

	\begin{Proof}
	If~\eref{3.7} holds, then substituting
\(
F(s;\gamma) = D^{-1}(\gamma)\circ \lindex[\mspace{-2mu}F]{}{\star}(s;\gamma)
\)
into~\eref{3.4}, we get~\eref{3.6}. Vice versa, if~\eref{3.6} is valid,
then from its comparison with~\eref{3.4} follows that
\(
D(\gamma) = \lindex[\mspace{-2mu}F]{}{\star}(t;\gamma)
				\circ \bigl(F(t;\gamma)\bigr)^{-1}
	  = \lindex[\mspace{-2mu}F]{}{\star}(s;\gamma)
				\circ \bigl(F(s;\gamma)\bigr)^{-1}
\)
is the required (independent of $s,t\in J$) isomorphism.
	\end{Proof}

	The above definition and results for linear transports along paths
deal with the general case concerning arbitrary vector bundles and are
therefore insensitive to the dimensionality of the bundle's base or fibres.
Below we point out some peculiarities of the case of a Hilbert bundle whose
fibres are generally infinitely dimensional.

	\begin{Defn}	\label{Defn3.2}
	A Hilbert fibre bundle is a fibre bundle whose fibres are
homeomorphic Hilbert spaces or, equivalently, whose (standard) fibre is a
Hilbert space.
	\end{Defn}

	In the present investigation we shall show that the Hilbert bundles
can be taken as a natural mathematical framework for a geometrical
formulation of quantum mechanics.
	For linear transports in a Hilbert bundle are valid all results
of~\cite{bp-normalF-LTP,bp-LTP-general, bp-TP-general} with a possible
exception of the ones in which (local) bases in the fibres are involved. The
cause for this is that the dimension of a Hilbert space is (generally)
infinity. So, there arise problems connected with the convergence or
divergence of the corresponding sums or integrals.  Below we try to avoid
these problems and to formulate our assertions and results in an invariant
way.

	Of course, propositions~\ref{Prop3.1} and~\ref{Prop3.2} remain valid
on Hilbert bundles; the only addition is that the vector spaces $V$ and
$\lindex[V]{}{\star}$ are now Hilbert spaces.

	Below, in Sect.~\ref{new-I} (see below the paragraph after
equation~\eref{4.12f}), we shall establish a result specific for the Hilbert
bundles that has no analogue in the general theory: a transport along paths is
Hermitian if and only if it is unitary. This assertion is implicitly contained
in~\cite[sect.~3]{bp-BQM-preliminary} (see the paragraph after equation~(3.6)
in it).

	In~\cite[sect.~3]{bp-normalF-LTP} are introduced the so-called
\emph{normal} frames for a linear transport along paths as a (local) field of
bases in which (on some set) the matrix of the transport is unit. Further in
this
series~\cite{bp-BQM-pictures+integrals}
we shall see that the normal frames realize the Heisenberg picture of motion
in the Hilbert bundle formulation of quantum mechanics.

\section[The Hilbert bundle description of quantum mechanics]
	{The Hilbert bundle description \\of quantum mechanics}
\label{new-I}
\setcounter{equation} {0}

	As we shall see in this investigation, the Hilbert bundles provide a
natural mathematical framework for a geometrical formulation of quantum
mechanics. In it all quantum\ndash mechanical quantities, such as
Hamiltonians, observables, wavefunctions, etc., have an adequate description.
For instance, the evolution of a systems is described as an appropriate
(parallel or, more precisely, linear) transport of system's state sections
along some path. We have to emphasize on the fact that the new bundle
formulation of quantum mechanics and the conventional one are completely
equivalent at the present stage.

	Before going on, we want to mention several works in which attempts
are made for a (partial) formulation of nonrelativistic quantum mechanics in
terms of fibre bundles.

	It seems that for the first time the real bundle approach to quantum
mechanics is developed in~\cite{Asorey&Carinena&Paramio} where the single
Hilbert space of quantum mechanics is replaced with an infinitely many copies
of it forming a bundle space over the 1\ndash dimensional `time' manifold
(\ie over $\mathbb{R}_+$). In this Hilbert fibre bundle the quantum evolution
is (equivalently) described as a kind of `parallel' transport of appropriate
objects over the bundle's base.

	Analogous construction, a Hilbert bundle over the system's phase
space, is used in the Prugove\v{c}ki's approach to quantum theory (see, e.g.
the references in~\cite{Drechsler/Tuckey-96}).

	In~\cite{Wilczek-Zee} is first mentioned about the gauge, \ie linear
connection, structure in quantum mechanics. That structure is pointed to be
connected with the system's Hamiltonian. This observation will find natural
explanation in our work.

	Some ideas concerning the interpretation of quantum evolution as a
kind of a `parallel' transport in a Hilbert bundle can also be found
in~\cite{Graudenz-94,Reuter}.

	After this introduction, we want to present some non-exactly rigorous
ideas and statements whose only purpose is the \emph{motivation} for
applying the fibre bundle formalism  to quantum mechanics. Another excellent
arguments and motives confirming this approach are given
in~\cite{Asorey&Carinena&Paramio}.

	Let $\base$ be a differentiable manifold, representing in our
context the space in which the (nonrelativistic) quantum-mechanical
objects `live', i.e.\  the usual 3-dimensional coordinate space
(isomorphic to $\mathbb{R}^3 $ with the corresponding structures).%
\footnote{\label{footnote-base}%
In the following $\base$ can naturally be considered also as the Minkowski
space-time of special relativity. In this case the below-defined
observer's trajectory $\gamma$ is his world line. But we avoid this
interpretation because only the nonrelativistic case is investigated
here. It is important to be noted that mathematically all of what
follows is valid in the case when by $\base$ is understood an arbitrary
differentiable manifold. The physical interpretation
of these cases will be given elsewhere.
}
Let $\gamma\colon J\to \base$, $J$ being an $\mathbb R$-interval, be
the trajectory of
an observer describing the behaviour of a quantum system at any moment
$t\in J$ by a state vector $\Psi_\gamma(t)$ depending on $t$ and, possibly,
on $\gamma$.%
\footnote{%
In this way we introduce the (possible) explicit dependence of the
description of a system's state on the concrete observer with respect to which
it is determined.%
}
For a fixed point $x=\gamma(t)\in \base$ the variety of state vectors
describing a quantum system and corresponding to different observers form a
Hilbert space $\bHil_{\gamma(t)}$ which
\emph{%
depends on $\gamma(t)=x$, but not on~$\gamma$ and~$t$ separately}.%
\footnote{
If there exists a global time, as in the nonrelativistic quantum
mechanics, the parameter $t\in J$ can be taken as such. Otherwise by
$t$ we have to understand the local (`proper' or `eigen-') time of a
concrete observer.%
}

	\begin{Rem}	\label{Rem-base}
	As we said above in footnote~\ref{footnote-base}, the next
considerations are completely valid mathematically if $\base$ is an arbitrary
differentiable manifold and $\gamma$ is a path in it. In this sense $\base$
and $\gamma$ are free parameters in our theory and their concrete choice is
subjected only to \emph{physical reasons}, first of all, ones requiring
adequate physical interpretation of the resulting theory. Typical candidates
for $\base$ are: the 3\ndash dimensional Euclidean space $\mathbb{E}^3$ or
$\mathbb{R}^3$, the 4\ndash dimensional Minkowski space $M^4$ of special
relativity or the Riemannian space $V_4$ of general relativity, the system's
configuration or phase space, the `time' manifold
$\mathbb{R}_+:=\{a:a\in\mathbb{R},\ a>0\}$, etc. Correspondingly, $\gamma$
obtains interpretation as particle's trajectory, its world line, and so on.
The degenerate case when $\base$ consists of a single point corresponds (up
to an isomorphism - see below) to the conventional quantum mechanics.
Throughout this work we most often take $\base=\mathbb{R}^3$ as a natural
choice corresponding to the non-relativistic case investigated here but, as we
said, this is not required by necessity. Elsewhere we shall see that
$\base=M^4$ or $\base=V_4$ are natural choices in the relativistic region.
An expanded commend on these problems will be given in
the concluding part of this series.
Here we want to note only that the interpretation of $\gamma$ as an
observer's (particle's) trajectory or world line, as accepted in this work,
is reasonable but not necessary one. May be more adequate is to interpret
$\gamma$ as a mean (in quantum-mechanical sense) trajectory of some point
particle but this does not change anything in the mathematical structure of
the bundle approach proposed here.
	\end{Rem}

	The spaces $\bHil_{\gamma(t)}$ must be isomorphic as, from physical
view-point, they simply represent the possible variety of state vectors
from different positions.
	In this way over $\base$ arises a natural bundle structure, viz.\ a
\emph{Hilbert bundle} $\HilB$ with a total space $\bHil$, projection
$\proj\colon \bHil\to \base$ and isomorphic fibres $\proj^{-1}(x):=\bHil_x$.
	Since $\bHil_x$, $x\in \base$ are isomorphic, there exists a Hilbert
space $\Hil$ and (linear) isomorphisms
$l_x\colon \bHil_x\to\Hil,\ x\in \base$.
Mathematically $\Hil$ is the typical (standard) fibre of $\HilB$.
(Note that we do not suppose local triviality, i.e.\  that for any $x\in \base$
there is a neighborhood $W\ni x$ in
$\base$ such that $\proj^{-1}(W)$ is homeomorphic to $\Hil\times W$.)
       The maps $\Psi_\gamma\colon J\to\proj^{-1}(\gamma(J))$ can be considered
as sections over any part of $\gamma$ without self-intersections
(see below).

	Now a natural question arises: how the quantum evolution in time in
the bundle constructed is described? There are two almost `evident' ways to do
this. On one hand, we can postulate the conventional quantum mechanics in
every fibre $\bHil_x$, \ie the Schr\"odinger equation for the state vector
$\Psi_\gamma(t)\in\bHil_{\gamma(t)}$ with $\bHil_{\gamma(t)}$ being (an
isomorphic copy of) the system's Hilbert space. But the only thing one gets
in this way is an isomorphic image of the usual quantum mechanics in any
fibre over $\base$. Therefore one can not expect some new results or
descriptions in this direction (see below~\eref{4.3b} and the comments after
it). On the other hand, we can demand the ordinary quantum mechanics to be
valid in the fibre $\Hil$ of the bundle $\HilB$. This means to identify
$\Hil$ with the system's Hilbert space of states and to describe the quantum
time evolution of the system via the vector
	\begin{equation}	\label{4.3}
\psi(t) = l_{\gamma(t)} ( {\Psi}_\gamma(t) ) \in \Hil
	\end{equation}
which evolves according to~(\ref{2.1}) or~(\ref{2.5}). This approach is
accepted in the present investigation. What we intend to do further, is, by
using the basic relation~\eref{4.3}, to `transfer' the quantum mechanics from
$\Hil$ to $\HilB$ or, in other words, to investigate the quantum evolution in
terms of the vector $\Psi_\gamma(t)$  connected with $\psi(t)$
via~\eref{4.3}. Since $l_x,\ x\in \base$ are isomorphisms, both descriptions
are \emph{completely equivalent}. This equivalence resolves a psychological
problem that may arise at first sight: the single Hilbert space $\Hil$ of
standard quantum theory~\cite{Dirac-PQM, Fock-FQM, Messiah-QM,
Prugovecki-QMinHS, L&L-3} is replaced with a, generally, infinite number
copies $\bHil_x$, $x\in\base$ thereof (cf.~\cite{Asorey&Carinena&Paramio}).
In the present investigation we shall show that the merit one gains from this
is an entirely geometrical reformulation of quantum mechanics in terms of
Hilbert fibre bundles.

	The above considerations were more or less heuristic ones.
The rigorous problem we want to investigate is the following. Let there be
given a quantum system described in the (nonrelativistic) quantum mechanics by
a state vector $\psi(t)$ satisfying the Schr\"odinger equation~\eref{2.5} and
belonging to the system's Hilbert space $\Hil$ of
states~\cite{Messiah-QM,L&L-3}. We \emph{postulate} that $\HilB$ is a Hilbert
fibre bundle with bundle space $\bHil$, base $\base$, projection
$\proj\colon\bHil\to\base$, and (typical, standard)
\emph{fibre coinciding with} $\Hil$. We suppose to be fixed a set of
isomorphisms $\{l_x :\ l_x\colon\bHil_x\to\Hil,\ x\in\base\}$
between the fibres $\bHil_x:=\proj^{-1}(x)$, $x\in\base$ and the typical
fibre $\Hil$. The base $\base$ is supposed to be a differentiable manifold
which, for definiteness, we shall identify with $\mathbb{R}^3$ (or with other
manifold `suitable' for the physical model; see remark~\ref{Rem-base}). Let
$\gamma\colon J\to\base$ be a path. In the case $\base=\mathbb{R}^3$ (resp.\
$\base=M^4,V_4$) we interpret $\gamma$ as a trajectory (resp.\ world line) of
an observer describing the behaviour of the quantum system under
consideration. If $\psi(t)$ is the complex vector-valued function of time
representing the system's state vector at a moment $t$, then our goal is to
describe the system's state at some instant of time $t$ via the vector
(cf.~\eref{4.3})
	\begin{equation}	\label{4.3a}
\Psi_\gamma(t) = l_{\gamma(t)}^{-1}(\psi(t))\in \bHil_{\gamma(t)}.
	\end{equation}
Since $l_x$, $x\in\base$ are isomorphisms, both descriptions of the quantum
evolution, through $\psi(t)$ and $\Psi_\gamma(t)$, are completely equivalent.

	Two important notes have to be made here. Firstly, the state
vectors in the bundle description generally explicitly depend on the
observer,\ie on the reference path $\gamma$,
which is depicted in the index $\gamma$ in $\Psi_\gamma(t)$.
This is on the contrary to the quantum mechanics where it is almost
everywhere implicitly assumed. And secondly, the bundle, as well as the
conventional, description of quantum mechanics is defined up to a linear
isomorphism(s). In fact, if $\imath\colon \Hil\to\Hil^\prime$,
$\Hil^\prime$ being a Hilbert space, is a linear isomorphism (which may
depend on the time $t$),
then $\psi^\prime(t)=\imath(\psi(t))$ equivalently describes
the evolution of the quantum system in $\Hil^\prime$.
	(Note that in this way, for $\Hil^\prime=\Hil$, one can obtain the
known pictures of motion in quantum mechanics --- see~\cite{Messiah-QM}%
.)
In the bundle case
the shift from $\Hil$ to $\Hil^\prime$ is described by the transformation
$l_x\to l_{x}^{\prime}:=\imath\circ l_x$ which reflects the
arbitrariness in the choice of the typical fibre (now $\Hil^\prime$ instead
of $\Hil$) of $\HilB$. There is also arbitrariness in the choice of the
fibres $\bHil_x=\proj^{-1}(x)$
which is of the same character as the one in the case of
$\Hil$, viz.\ if $\imath_x\colon \bHil_x\to \bHil_{x}^{\prime},\ x\in \base$
are linear
isomorphisms, then the fibre bundle $(\bHil^\prime,\proj^\prime,\base)$ with
$\bHil^\prime:=\bigcup_{x\in \base}\bHil_{x}^{\prime},\
\left.\proj^\prime\right|_{\bHil_{x}^{\prime}}:= \proj\circ\imath_{x}^{-1}$,
typical fibre $\Hil$, and isomorphisms
$l^\prime_x := l_x\circ\imath_{x}^{-1}$
can equivalently be used to describe the evolution of a quantum
system. In the most general case, we have a fibre bundle
$(\bHil^\prime,\proj^\prime,\base)$ with fibres
$\bHil_{x}^{\prime}=\imath_{x}^{-1}(\bHil_x)$, typical fibre
$\Hil^\prime = \imath(\Hil)$,
and isomorphisms
\(
l_x^\prime :=
\imath\circ l_x\circ\imath_{x}^{-1}\colon \bHil_{x}^{\prime}\to\Hil^\prime.
\)
Further we will not be interested in such generalizations. Thus, we shall
suppose that all of the mentioned isomorphisms are fixed in such a way that
the evolution of a quantum system will be described in a fibre bundle
$\HilB$ with fixed isomorphisms $\{l_x,\ x\in \base\}$ such that
$l_x\colon \bHil_x\to\Hil$, where $\Hil$ is the Hilbert space in which the
system's
evolution is described through the usual Schr\"odinger picture of motion.%
\footnote{%
Note that in the mentioned context the Schr\"odinger picture of motion plays
the same r\^ole as the inertial frames in the Newtonian mechanics.%
}

	So, in the Schr\"odinger picture a quantum system is described by a
state vector $\psi$ in $\Hil$.%
\footnote{%
The concrete choice of $\Hil$ is insignificant for the following, the only
important thing is the fulfillment of the Schr\"odinger equation for the
evolving vectors in it.%
}
	Generally~\cite{Neumann-MFQM}
$\psi$ depends (maybe implicitly) on the observer with
respect to which the evolution is studied%
\footnote{%
Usually this dependence is not written explicitly, but it is always presented
as actually $t$ is the time with respect to a given observer.
}
	and it satisfies the Schr\"odinger
equation~(\ref{2.5}). We shall refer to this representation of quantum
mechanics as a
\emph{Hilbert space description}.
In the new
\emph{(Hilbert fibre) bundle description},
which will be studied below, the linear isomorphisms
$l_x\colon \bHil_x=\proj^{-1}(x)\to\Hil,\ x\in \base$ are supposed
arbitrarily fixed%
\footnote{%
The particular choice of $\{l_x\}$  (and, consequently, of the fibres
$\bHil_x$)
is inessential for our investigation.%
}
and the quantum systems are described by a
\emph{state section along paths}
$\Psi$ of a fibre bundle $\HilB$ whose typical fibre is the
Hilbert space $\Hil$ (the same Hilbert space
as in the Hilbert space description).

	Here the term (state) section along paths needs some
explanations and correct definition. The proper bundle analogue of
$\psi(t)\in\Hil$ is
$\Psi_\gamma(t)\in \bHil_{\gamma(t)}$, given by~(\ref{4.3a}),
which explicitly
depends on the observer's trajectory (world line in the special relativity
interpretation). Let $J^\prime\subseteq J$ be any subinterval of $J$ on which
$\gamma$ is without self-intersections, i.e.\  if $s,t\in J^\prime$ and
$s\not=t$, then $\gamma(s)\neq\gamma(t)$.
The map
\(
\Psi_{\gamma|J^\prime}\colon \gamma(J^\prime)\to\proj^{-1}(\gamma(J^\prime))
\subset \bHil
\)
given by
\(
\Psi_{\gamma|J^\prime} \colon x\mapsto\Psi_\gamma(t),\ x\in\gamma(J^\prime),
\)
for the unique $t\in J^\prime$ for which $\gamma(t)=x$,
is a depending on $\gamma$ section of the restricted bundle
$\left.\HilB\right|_{\gamma(J^\prime)}$, i.e.\
\(
\Psi_{\gamma|J^\prime}\in\Sec
\left(  \left.\HilB\right|_{\gamma(J^\prime)}  \right).
\)%
\footnote{%
Since in the special relativity interpretation $\gamma$ is observer's
world line, the path $\gamma$ can not have self-intersections (for real
particles and (extended) bodies). In this case the map $\Psi_\gamma$ is
a section over the whole set $\gamma(J)$.%
}
	Generally we can put \(
\Psi_\gamma\colon x\mapsto\{\Psi_\gamma(t):\quad t\in J,\ \gamma(t)=x\}
\)
for every $x\in \base$. Evidently
$\Psi_\gamma\colon x\mapsto\emptyset$, $\emptyset$ being the empty set, for
$x\not\in\gamma(J),$
$\proj\circ\left.\Psi_\gamma\right|_{\gamma(J)} = \id_{\gamma(J)}$,
and at the points of self-intersection of $\gamma$, if any, $\Psi_\gamma$ is
multiple valued, with the number of its values being equal to one plus the
number of self-intersections of $\gamma$ at the corresponding point.
	We call \emph{section along paths} any map
$\Psi\colon\gamma\mapsto\Psi_\gamma$, where
$\Psi_\gamma\colon \base\to \bHil$
may be multiple valued and such that
$\proj\circ\left.\Psi_\gamma\right|_{\gamma(J)}=\id_{\gamma(J)}$ and
$\Psi_\gamma\colon x\mapsto\emptyset$ for $x\not\in\gamma(J)$.
So, the above-defined object $\Psi_\gamma$ is a section along $\gamma$.
It is single valued, and consequently a section over $\gamma(J)$ in the
usual sense~\cite{Husemoller}, iff $\gamma$ is without self-intersections.

	We want also to mention explicitly the natural interpretation of
$\Psi$ as a \emph{lifting of paths}, which is
suggested by the notation used (see, e.g.,~\eref{4.3a}). Actually
(cf.~[chapter~I, sect.~16 and chapter~III, sect.~7]\cite{Sze-Tsen}), a
lifting of paths (from $\base$ to $\bHil$) is a map
$\Psi\colon\gamma\mapsto\Psi_\gamma$ assigning to any path
$\gamma\colon J\to\base$ a path $\Psi_\gamma\colon J\to\bHil$, lifting of
$\gamma$ (from $\base$ to $\bHil$), such that
$\proj\circ\Psi_\gamma:=\gamma$. Evidently, the map
$\Psi_\gamma\colon t\mapsto\Psi_\gamma(t)$ given by~\eref{4.3a} is a lifting
of $\gamma$; therefore $\Psi\colon\gamma\mapsto\Psi_\gamma$ is lifting of
$\gamma$ which is single\ndash valued irrespectively of the existence of
self\ndash intersections of $\gamma$.

	Generally, to any vector $\varphi\in\Hil$ there corresponds a
unique (global) section $\overline{\Phi}\in\Sec\HilB$ defined via
	\begin{equation}	\label{4.3b}
\overline{\Phi}\colon x\mapsto\overline{\Phi}_x :=
			l_{x}^{-1}(\varphi)\in \bHil_x,
\qquad x\in \base,\quad \varphi\in\Hil.
	\end{equation}
Consequently to a state vector $\psi(t)\in\Hil$ one can assign the (global)
section $\overline{\Psi}(t)$,
\(
\overline{\Psi}(t)\colon x\mapsto\overline{\Psi}_x(t) =
					l_{x}^{-1}(\psi(t))\in \bHil_x
\)
and thus obtaining in $\bHil_x$ for every $x\in \base$ an isomorphic picture
of (the evolution in) $\Hil$. But in this way one can not obtain something
significantly new as the evolution in $\Hil$ is simply replaced with the
(linearly isomorphic to it) evolution in $\bHil_x$ for any arbitrary fixed
$x\in \base$. This reflects the above-mentioned fact that the quantum
mechanical description is defined up to linear isomorphism(s).
Besides, on the contrary to the bundle description, in this way one
looses the explicit dependence on the observer. So in it one
can't get something really new with respect to the Hilbert space description.
\vspace{2.3ex}

	Below we are going to define some structures and maps specific to
Hilbert bundles and having a relation to the Hilbert bundle description of
quantum mechanics.

	Denote by $\langle\cdot | \cdot\rangle_x$ the Hermitian scalar
product in $\bHil_x$. We demand the isomorphisms $l_x$ to preserve not only
the linear but also the metric structure of the bundle, i.e.\
\(
\langle \varphi | \psi \rangle =
\langle l_{x}^{-1}\varphi|l_{x}^{-1}\psi \rangle _x,\
\varphi,\psi\in\Hil.
\)
Consequently $l_x$ transform the metric structure from $\Hil$ to $\bHil_x$
for every $x\in \base$ according to
	\begin{equation}	\label{4.8}
\langle \cdot|\cdot \rangle _x=
\langle l_{x}\cdot|l_{x}\cdot \rangle, \qquad x\in \base
	\end{equation}
and, consequently, from $\bHil_x$ to $\Hil$ through
	\begin{equation}	\label{4.8a}
\langle \cdot|\cdot \rangle =
\langle l_{x}^{-1}\cdot|l_{x}^{-1}\cdot \rangle _x, \qquad x\in \base.
	\end{equation}

	Defining the
\emph{Hermitian conjugate}
map (operator)
\( \mor{A}_{x}^{\ddag}\colon \Hil\to \bHil_x \) of a map
\( \mor{A}_x\colon  \bHil_x \to \Hil \) by
	\begin{equation}	\label{4.9}
\langle \mor{A}_{x}^{\ddag}\varphi|\chi_x \rangle_x :=
\langle \varphi | \mor{A}_x\chi_x\rangle,
\qquad \varphi\in\Hil,\quad \chi_x\in \bHil_x,
	\end{equation}
we find (see~(\ref{4.8}))
	\begin{equation}	\label{4.10}
\mor{A}_{x}^{\ddag} =
l_{x}^{-1}\circ \left( \mor{A}_x\circ l_{x}^{-1} \right)^\dag
	\end{equation}
where the dagger denotes Hermitian conjugation in $\Hil$ (see~(\ref{2.4})).

	We call a map $\mor{A}_x$ \emph{unitary} if
	\begin{equation}	\label{4.10a}
\mor{A}_{x}^{\ddag} = \mor{A}_{x}^{-1} .
	\end{equation}
Evidently, the isomorphisms $l_x$ are unitary in this sense:
	\begin{equation}	\label{4.10b}
l_{x}^{\ddag} = l_{x}^{-1}.
	\end{equation}

	Similarly, the
\emph{Hermitian conjugate}
map to a map
$\mor{A}_{x\to y}\in\{\mor{C}_{x\to y}\colon\bHil_x\to\bHil_y,\ x,y\in\base\}$
is a map
\( \mor{A}_{x\to y}^{\ddag}\colon \bHil_x\to \bHil_y \)
defined via
	\begin{equation}	\label{4.11}
\langle \mor{A}_{x\to y}^{\ddag} \Phi_x| \Psi_y \rangle_y :=
\langle \Phi_x| \mor{A}_{y\to x}\Psi_y \rangle_x,
\qquad \Phi_x\in \bHil_x,\quad \Psi_y\in \bHil_y.
	\end{equation}
	Its explicit form is
	\begin{equation}	\label{4.12}
\mor{A}_{x\to y}^{\ddag} =
l_{y}^{-1}\circ
\left( l_x\circ \mor{A}_{y\to x}\circ l_{y}^{-1} \right)^\dag
\circ l_x.
	\end{equation}
As $(\ope{A}^\dag)^\dag\equiv \ope{A}$ for any
$\ope{A}\colon \Hil\to\Hil$, we have
	\begin{equation}	\label{4.12a}
\left( \mor{A}_{x\to y}^{\ddag} \right)^\ddag = \mor{A}_{x\to y} .
	\end{equation}

	 If
\(
\mor{B}_{x\to y}\in\{\mor{C}_{x\to y}\colon\bHil_x\to\bHil_y,\ x,y\in \base\}
\),
then a simple verification shows
	\begin{equation}	\label{4.12b}
\left(\mor{B}_{y\to z}\circ \mor{A}_{x\to y} \right)^\ddag =
\mor{A}_{y\to z}^\ddag\circ\mor{B}_{x\to y}^\ddag, \qquad x,y,z\in \base.
	\end{equation}

	A map $\mor{A}_{x\to y}$ is called
\emph{Hermitian}
if
	\begin{equation}	\label{4.12c}
\mor{A}_{x\to y}^\ddag = \mor{A}_{x\to y}.
	\end{equation}
A simple calculation proves that the maps
\( l_{x\to y}:=l_y^{-1}\circ l_x \)
are Hermitian.

	A map \(\mor{A}_{x\to y}\) is called
\emph{unitary}
if it has a left inverse map and
	\begin{equation}	\label{4.12d}
\mor{A}_{x\to y}^\ddag = \mor{A}_{y\to x}^{-1},
	\end{equation}
where
\(
\mor{A}_{x\to y}^{-1}\colon \bHil_y\to\bHil_x
\)
is the \emph{left} inverse of
\(\mor{A}_{x\to y}\),
i.e.\
\(
\mor{A}_{x\to y}^{-1}\circ \mor{A}_{x\to y} := \id_{\bHil_x}
\).

	A simple verification by means of~\eref{4.11} shows the equivalence
of~\eref{4.12d} with
	\begin{equation}	\tag{\ref{4.12d}$^\prime$}	\label{4.12d'}
\langle\mor{A}_{y\to s}\cdot | \mor{A}_{y\to s}\cdot\rangle _x
   = \langle\cdot | \cdot\rangle _y
   \colon\bHil_y\times\bHil_y\to\mathbb{C},
	\end{equation}
\ie the unitary maps are fibre-metric compatible in a sense that they
preserve the fibre scalar (inner) product. Such maps will be called
\emph{fibre\ndash isometric} or simply \emph{isometric}.

	It is almost evident that the maps $l_{x\to y}=l_{y}^{-1}\circ l_x$
are unitary, that is we have:%
\footnote{%
The Hermiticity and at the same time unitarity of $l_{x\to y}$ is not
incidental as they define a (flat) linear transport (along paths or along the
identity map of $M$) in $\HilB$ (see~(\ref{3.4}) and below the paragraph
after~(\ref{4.12f})).%
}
	\begin{equation}	\label{4.12h}
l_{x\to y}^{\ddag}=l_{x\to y}=l_{y\to x}^{-1},
\qquad
l_{x\to y}:=l_{y}^{-1}\circ l_x  \colon\proj^{-1}(x)\to\proj^{-1}(y) .
	\end{equation}

	We call a (possibly linear) transport along paths in $\HilB$
\emph{Hermitian} or \emph{unitary} if it satisfies
respectively~(\ref{4.12c}) or~(\ref{4.12d}) in which $x$, and $y$ are
replaced with arbitrary values of the parameter of the transportation
path, i.e.\  if respectively
	\begin{align}	\label{4.12e}
\left( L_{s\to t}^{\gamma} \right)^\ddag &= L_{s\to t}^{\gamma} ,
\qquad s,t\in J,\quad \gamma\colon J\to \base,
\\	\label{4.12f}
\left( L_{s\to t}^{\gamma} \right)^\ddag &=
				\left( L_{t\to s}^{\gamma} \right)^{-1}.
	\end{align}

A simple corollary from~(\ref{3.3}) is the equivalence of ~(\ref{4.12e})
and ~(\ref{4.12f}); therefore, a
\emph{%
transport along paths in a Hilbert bundle is Hermitian
if and only if it is unitary%
},
i.e.\  these concepts are equivalent. For such transports we say that they are
\emph{consistent} or \emph{compatible} with the Hermitian structure (metric
(inner product)) of the Hilbert bundle~\cite{bp-TP-morphisms}.
Evidently, they are \emph{isometric} fibre maps along the paths they act.
Therefore, a transport along paths in a Hilbert bundle is isometric iff it is
Hermitian of iff it is unitary.%
\footnote{%
The author thanks prof. James Stasheff (Math-UNC, Chapel Hill, NC, USA) for
suggesting in July 1998 the term ``isometric transport'' in the context
given.%
}

	Let $\mor{A}$ be a bundle morphism of $\HilB$, i.e.\
$\mor{A}\colon \bHil\to \bHil$ and $\proj\circ \mor{A} = \id_\base$,
and $\mor{A}_x:=\left.\mor{A}\right|_{\bHil_x}$.
The \emph{Hermitian conjugate}
bundle morphism $\mor{A}^\ddag$ to $\mor{A}$ is defined by (cf.~(\ref{4.11}))
	\begin{equation}	\label{4.morphism}
\langle \mor{A}^\ddag\Phi_x | \Psi_x \rangle_x :=
\langle \Phi_x | \mor{A}\Psi_x \rangle_x,
\qquad \Phi_x,\Psi_x\in \bHil_x.
	\end{equation}
Thus (cf.~(\ref{4.12}))
	\begin{equation}	\label{4.15}
\mor{A}_{x}^{\ddag} := \left.\mor{A}^\ddag\right|_{\bHil_x} =
l_{x}^{-1}\circ\left(l_x\circ \mor{A}_x\circ l_{x}^{-1}\right)^\dag\circ l_x.
	\end{equation}

	A bundle morphism $\mor{A}$ is called
\emph{Hermitian}
if \(\mor{A}_{x}^{\ddag}=\mor{A}_x\) for every $\ x\in \base$, i.e.\  if
	\begin{equation}	\label{4.16}
\mor{A}^\ddag=\mor{A},
	\end{equation}
and it is called
\emph{unitary}
if \(\mor{A}_{x}^{\ddag}=\mor{A}_x^{-1}\) for every $\ x\in \base$, i.e.\  if
	\begin{equation}	\label{4.17}
\mor{A}^\ddag=\mor{A}^{-1}.
	\end{equation}
Using~\eref{4.morphism}, we can establish the equivalence of~\eref{4.17} and
	\begin{equation}	\tag{\ref{4.17}$^\prime$}	\label{4.17'}
\langle\mor{A}\cdot | \cdot\mor{A}\rangle_x = \langle\cdot | \cdot\rangle_x
   \colon\bHil_x\times\bHil_x\to\mathbb{C}.
	\end{equation}
Consequently the unitary morphisms are fibre-metric compatible, \ie they are
\emph{isometric} in a sense that they preserve the fibre Hermitian scalar
(inner) product.

\section {The (bundle) evolution transport}
\label{IV}
\setcounter{equation} {0}

	Using~(\ref{2.1}), we get
\(
\psi(t_3)=\ope{U}(t_3,t_2)\psi(t_2) =
\ope{U}(t_3,t_2) [ \ope{U}(t_2,t_1)\psi(t_1) ],\
\psi(t_3)=\ope{U}(t_3,t_1)\psi(t_1),\ \mathrm{and}\
\psi(t_1)=\ope{U}(t_1,t_1)\psi(t_1)
\)
for every moments $t_1,t_2,t_3$ and arbitrary state vector $\psi$. Hence
	\begin{align}
\ope{U}(t_3,t_1) &= \ope{U}(t_3,t_2)  \circ \ope{U}(t_2,t_1), \label{4.1} \\
\ope{U}(t_1,t_1) &= \id_{\Hil}.			\label{4.2}
\\
\intertext{Besides, by definition,
$\ope{U}(t_2,t_1)\colon \Hil\to\Hil$ is a linear unitary operator, i.e.\  for
\( \lambda_i\in\mathbb{C} \) and
\( \psi_i(t_1)\in\Hil \), $i=1,2$,
we have:}
\ope{U}(t_2,t_1) \biggl(\sum_{i=1,2}^{} \lambda_i\psi_i(t_1)\biggr) &=
\sum_{i=1,2}^{}\lambda_i \ope{U}(t_2,t_1) \psi_i(t_1),	\label{4.2a} \\
\ope{U}^\dag(t_1,t_2) &= \ope{U}^{-1}(t_2,t_1). 		\label{4.2b}
\\
\intertext{From~(\ref{4.1}) and~(\ref{4.2}), evidently, follows}
\ope{U}^{-1}(t_2,t_1) &= \ope{U}(t_1,t_2)		\label{4.2c}
\intertext{and consequently}
\ope{U}^\dag(t_1,t_2) &= \ope{U}(t_1,t_2).		\label{4.2d}
	\end{align}

	If one takes as a primary object the Hamiltonian $\Ham$, then
these facts are direct consequences of~(\ref{2.8}).

	Thus the properties of the evolution operator are very similar
to the ones defining a ((flat) Hermitian) linear transport along paths
in a Hilbert bundle. In fact, below we show that the evolution operator
is a kind of such transport. (Note that this description is not unique.)

	The bundle analogue of the evolution operator
$\ope{U}(t,s)\colon \Hil\to\Hil$ is a linear operator
\(
\mor{U}_\gamma(t,s)\colon \bHil_{\gamma(s)} \to \bHil_{\gamma(t)},\ s,t\in J
\)
such that
	\begin{equation}	\label{4.4}
\Psi_\gamma(t) = \mor{U}_\gamma(t,s) \Psi_\gamma(s)
	\end{equation}
for every instants of time $s,t\in J$. Analogously to~(\ref{4.1})
and~(\ref{4.2}), now we have:
	\begin{align}				\label{4.5}
\mor{U}_\gamma(t_3,t_1) &= \mor{U}_\gamma(t_3,t_2)\circ
\mor{U}_\gamma(t_2,t_1), \qquad t_1,t_2,t_3\in J,	\\ \label{4.6}
\mor{U}_\gamma(t,t) &= \id_{\bHil_{\gamma(t)}}, \qquad t\in J.
	\end{align}
	We call $\mor{U}$
\emph{bundle evolution operator} or \emph{evolution transport} (see below).

	Comparing~(\ref{4.4}) with~(\ref{2.1}) and using~(\ref{4.3a}),
we find
	\begin{alignat}{2}	\label{4.7}
\mor{U}_\gamma(t,s) &=
		l_{\gamma(t)}^{-1}\circ \ope{U}(t,s) \circ l_{\gamma(s)},
&\qquad s,t&\in J
\\ \intertext{or}
			\label{4.7'}
\ope{U}(t,s)  &=
l_{\gamma(t)} \circ \mor{U}_\gamma(t,s) \circ l_{\gamma(s)}^{-1},
&\qquad s,t&\in J.
	\end{alignat}
This shows the equivalence of the description of evolution
of quantum systems via $\ope{U}$ and~$\mor{U}_\gamma$.

	A trivial corollary of~(\ref{4.7}) is the \emph{linearity} of
$\mor{U}_\gamma$ and
	\begin{equation}	\label{4.7a}
\mor{U}_\gamma^{-1}(t,s) = \mor{U}_\gamma(s,t) .
	\end{equation}

	As $l_x\colon \bHil_x\to\Hil,\ x\in \base$ are linear isomorphisms,
from~(\ref{4.5})--(\ref{4.7}) follows that
\(
\mor{U}\colon \gamma\mapsto
\mor{U}_\gamma\colon (s,t)\mapsto \mor{U}_\gamma(s,t)
=:\mor{U}_{t\to s}^{\gamma} \colon  \bHil_{\gamma(t)} \to \bHil_{\gamma(s)}
\)
is a linear transport along paths in $\HilB$.%
\footnote{%
\label{L-transport:evolution-transport}%
In the context of quantum mechanics it is more natural to define
$\mor{U}_\gamma(s,t)$ from $\bHil_{\gamma(t)}$ into $\bHil_{\gamma(s)}$
instead from
$\bHil_{\gamma(s)}$ into $\bHil_{\gamma(t)}$, as is the map
\(
\mor{U}_{s\to t}^{\gamma}=\mor{U}_\gamma(t,s) \colon
		\bHil_{\gamma(s)} \to \bHil_{\gamma(t)}
\).
The latter notation is better in the general theory of transports along
paths~\cite{bp-normalF-LTP,bp-LTP-general}. Consequently, when applying
results from~\cite{bp-normalF-LTP,bp-LTP-general}, we have to remember that
they are valid for the maps $\mor{U}_{s\to t}^{\gamma}$
(or $\mor{U}^\gamma\colon (s,t)\mapsto \mor{U}_{s\to t}^{\gamma}$).
That is why for the usage of some results  concerning general linear
transports along paths from~\cite{bp-normalF-LTP,bp-LTP-general}
for $\mor{U}_\gamma(s,t)$ or $\mor{U}_\gamma$ one has to write them for
$\mor{U}_{s\to t}^{\gamma}$ (or $\mor{U}^\gamma$) and then to use the
connection
\(
\mor{U}_{s\to t}^{\gamma} = \mor{U}_\gamma(t,s)=\mor{U}_{\gamma}^{-1}(s,t)
\)
(or $\mor{U}^\gamma=\mor{U}_{\gamma}^{-1}$).
Some results for $\mor{U}_{s\to t}^{\gamma}$ and $\mor{U}_\gamma(s,t)$
coincide but
this is not always the case. In short, the results for linear
transports along paths are transferred to the considered in this work
case by replacing $L_{s\to t}^{\gamma}$ with
$\mor{U}_\gamma(t,s)=\mor{U}_{\gamma}^{-1}(s,t)$.%
}
 This transport is
\emph{Hermitian} (see Sect.~\ref{new-I}).
	In fact, applying~(\ref{4.12}) to $\mor{U}_\gamma(t,s)$ and
using~(\ref{4.7}), we get
	\begin{equation}	\label{4.13}
\mor{U}_{\gamma}^{\ddag}(t,s) =
l_{\gamma(t)}^{-1} \circ \ope{U}^\dag(s,t) \circ l_{\gamma(t)}.
	\end{equation}
So, using~(\ref{4.2d}), once again~(\ref{4.7}), and~(\ref{4.2c}), we find
	\begin{equation}	\label{4.14}
\mor{U}_{\gamma}^{\ddag}(t,s) = \mor{U}_\gamma(t,s) =
				\mor{U}_{\gamma}^{-1}(s,t).
	\end{equation}

	Hence $\mor{U}_\gamma(t,s)$ is simultaneously Hermitian and unitary
operator, as it should be for any Hermitian or unitary transport along paths
in a Hilbert bundle (see Sect.~\ref{new-I}). Consequently, the evolution
transport is an isometric transport along paths.

	In this way, we see that the bundle evolution operator $\mor{U}$ is a
Hermitian (and hence unitary) linear transport along paths in
$\HilB$. Consequently, to any unitary evolution operator $\ope{U}$
in the Hilbert space $\Hil$ there corresponds a unique isometric linear
transport $\mor{U}$ along paths in the Hilbert bundle $\HilB$ and vice versa.

\section{Conclusion}
\label{conclusion-I}

	In the present work we have prepared the background for a full
self-consistent fibre bundle formulation of nonrelativistic quantum mechanics.
For this purpose we constructed a Hilbert (vector) fibre bundle replacing
now the conventional Hilbert space of quantum mechanics. On this scene, as was
shown here, the conventional quantum evolution is described by a suitable
linear transport along paths.

	An advantage of the bundle description of quantum mechanics is that
it does not make use of any particular model of the base $\base$. But on
this model depends the interpretation of `time' $t$ used. For instance, if we
take $\base$ to be the 3-dimensional Euclidean space $\mathbb{E}^3$ of
classical (or quantum) mechanics, then $t$ is natural to be identified with
the absolute Newtonian (global) time. However, if $\base$ is taken to be the
Minkowski 4-dimensional space $M_4$, then it is preferable to take $t$ to be
the proper time of some (local) observer, but the global coordinate time in
some frame can also play the r\^ole of $t$. Principally different is the
situation when the pseudo-Riemannian space $V_4$ of general relativity is
taken as $\base$: now $t$ \emph{must} be the local time of some observer as a
global time does not generically exist.

	Generally, the space-time model $\base$ is external to (bundle)
quantum mechanics and has to be determined by another theory, such as special
or general relativity. This points to a possible field of research: a
connection between the quantities of the total bundle space with a concrete
model of $\base$ may result in a completely new theory. Elsewhere we shall
show that just this is the case with relativistic quantum mechanics.

	The development of the bundle approach to quantum mechanics will be
done in the continuation of this paper. In particular, we intend to
investigate the following topics from the novel fibre bundle view-point:
equations of motion, description of observables, pictures and integrals of
motion, mixed states, interpretation of the theory and possible ways for its
further development and generalizations.


\addcontentsline{toc}{section}{References}

\bibliography{bozhopub,bozhoref}
\bibliographystyle{unsrt}

\addcontentsline{toc}{subsubsection}{\vspace{1ex}This article ends at page}

\end{document}